\documentclass[aps,prb,showpacs,longbibliography,superscriptaddress,citeonline,twocolumn]{revtex4-2}
\usepackage{amsmath,amssymb,bm}
\usepackage{graphicx}
\usepackage{xspace}
\usepackage{latexsym}
\usepackage{xcolor}
\definecolor{LinkColor}{rgb}{0.256,0.439,0.588}
\usepackage{hyperref}
\hypersetup{
colorlinks=true,
citecolor=LinkColor,
linkcolor=LinkColor,
urlcolor=LinkColor
}
\usepackage{orcidlink}

\newcommand{\scbo}{SrCu$_{2}$(BO$_{3}$)$_2$~}

\begin{document}

\title{Pinwheel-shaped bound triplet pairs in the magnetization plateaus of SrCu$_2$(BO$_3$)$_2$}

\author{Igor Vinograd\,\orcidlink{0000-0002-1054-3741}}
\thanks{These authors contributed equally to this work.}
\affiliation{
Laboratoire National des Champs Magn\'{e}tiques Intenses, LNCMI-CNRS (UPR3228), EMFL, \\
Univ. Grenoble Alpes, Univ. Toulouse, INSA-T, 38042 Grenoble Cedex 9, France}

\author{Philippe Corboz\,\orcidlink{0000-0003-1024-1230}}
\altaffiliation{These authors contributed equally to this work.}
\affiliation{
Institute for Theoretical Physics,
University of Amsterdam, Science Park 904, 1098 XH Amsterdam, The Netherlands}

\author{Steffen Kr\"{a}mer\,\orcidlink{0000-0002-6107-3583}}
\affiliation{
Laboratoire National des Champs Magn\'{e}tiques Intenses, LNCMI-CNRS (UPR3228), EMFL, \\
Univ. Grenoble Alpes, Univ. Toulouse, INSA-T, 38042 Grenoble Cedex 9, France}

\author{Yanan Li\,\orcidlink{0000-0003-3851-9425}}
\affiliation{
Laboratoire National des Champs Magn\'{e}tiques Intenses, LNCMI-CNRS (UPR3228), EMFL, \\
Univ. Grenoble Alpes, Univ. Toulouse, INSA-T, 38042 Grenoble Cedex 9, France}
\affiliation{
Present address: College of Electronic and Information Engineering, Liaoning Technical University,
125105, Huludao, Liaoning, China}

\author{Fr\'ed\'eric Mila\,\orcidlink{0000-0003-4306-7996}}
\affiliation{
Institute of Physics, \'{E}cole Polytechnique F\'{e}d\'{e}rale de Lausanne (EPFL), CH-1015 Lausanne, Switzerland}

\author{Hiroshi Kageyama\,\orcidlink{0000-0002-3911-9864}}
\affiliation{
Graduate School of Engineering, Kyoto University, Nishikyo-ku, Kyoto 615-8510, Japan}

\author{Masashi Takigawa\,\orcidlink{0000-0002-4516-5074}}
\affiliation{
Professor Emeritus, Institute for Solid State Physics, University of Tokyo, 5-1-5 Kashiwanoha, Kashiwa, Chiba 277-8581, Japan}

\author{Mladen Horvati{\'c}\,\orcidlink{0000-0001-7161-0488}}
\email{mladen.horvatic@lncmi.cnrs.fr}
\affiliation{
Laboratoire National des Champs Magn\'{e}tiques Intenses, LNCMI-CNRS (UPR3228), EMFL, \\
Univ. Grenoble Alpes, Univ. Toulouse, INSA-T, 38042 Grenoble Cedex 9, France}
\date{\today}

\begin{abstract}
The sequence of magnetization plateaus at 1/8, 2/15, 1/6, 1/4, 1/3, 2/5, and 1/2 of the saturation observed in SrCu$_2$(BO$_3$)$_2$ remained a puzzle until tensor-networks-based numerical simulations suggested that the low-magnetization plateaus are stabilized as Wigner crystals of spin-2 bound states, and not as one of the standard configurations (semiclassical up-down, or crystals of triplets). We report on $^{63,65}$Cu nuclear magnetic resonance spectra up to 41.9~T: Based on constraints deduced from the 1/3 plateau, we show that the spectra in the 1/8 and 2/15 plateaus indeed agree with the prediction for the spin-2 bound states, while being incompatible with a crystal of triplets. This adds the formation of Wigner crystals of bound states as an alternative fundamental paradigm in the theory of magnetization plateaus.

\end{abstract}
\maketitle

\section{Introduction}
Fractional magnetization plateaus are ubiquitous in frustrated quantum magnets \cite{Book2011}. They correspond to incompressible phases in which an increasing magnetic field is unable to increase the magnetization. The microscopic origin is not unique, but most plateaus observed experimentally or numerically have received a simple explanation in terms of one of the following three paradigms: (i) the stabilization of a classical configuration with up and down spins by quantum fluctuations, as in the 1/3 plateau of the triangular lattice~\cite{Nishimori1986,Chubukov1991}, (ii) a crystal of triplets in a sea of singlets in dimer-based antiferromagnets, as in the 1/2 plateau of the fully frustrated ladder~\cite{Mila1998,Okazaki2000} or in the Shastry-Sutherland lattice discussed here ~\cite{Shastry1981, Kageyama1999, Onizuka00,Kodama2002a}, or (iii) a crystal of localized magnons in a polarized state, as in the 7/9 plateau of the kagome lattice~\cite{Schulenburg2002,Capponi2013}. However, some low-magnetization plateaus, such as the 1/9 plateau of the kagome model~\cite{Nishimoto2013,He2024} or the 2/15 plateau observed in the Shastry-Sutherland compound \scbo\cite{Takigawa2013} could not be explained along these lines.

Thanks to the development of tensor-network based methods~\cite{verstraete2004,jordan2008}, and to a systematic investigation of the spin-1/2 Shastry-Sutherland model~\cite{Shastry1981} in a magnetic field, allowing for very large unit cells, the situation has recently improved. It has been suggested that in the sequence of plateaus at 1/8, 2/15, 1/6, 1/4, 1/3, 2/5 and 1/2 of the magnetization in \scbo \cite{Takigawa2013,Jaime2012,Matsuda2013,Shi2022}, the mechanism behind the first three plateaus relies on the formation of spin-2 bound states~\cite{Momoi2000} that form Wigner crystals with large unit cells~\cite{Corboz2014}. This theory is the only one able to explain the observed low-magnetization sequence of plateaus, but it could not be confirmed or denied so far because of the lack of experiments able to distinguish between these phases and standard Wigner crystals of triplets. In the present paper, we report on $^{63,65}$Cu nuclear magnetic resonance (NMR) experiments at fields up to 41.9 T that clearly confirm the formation of Wigner crystals of spin-2 bound states in the 1/8 and 2/15 plateaus. This conclusion relies on a precise determination of the hyperfine fields thanks to the 41.9 T NMR spectrum in the non-controversial 1/3 plateau, and on a direct comparison of the NMR spectra and numerical histograms of spin polarizations in the 1/8 and 2/15 plateaus, made possible by this new determination of the hyperfine fields.


\begin{figure*}[t!]
    \centering
    \includegraphics[width=1.00\textwidth]{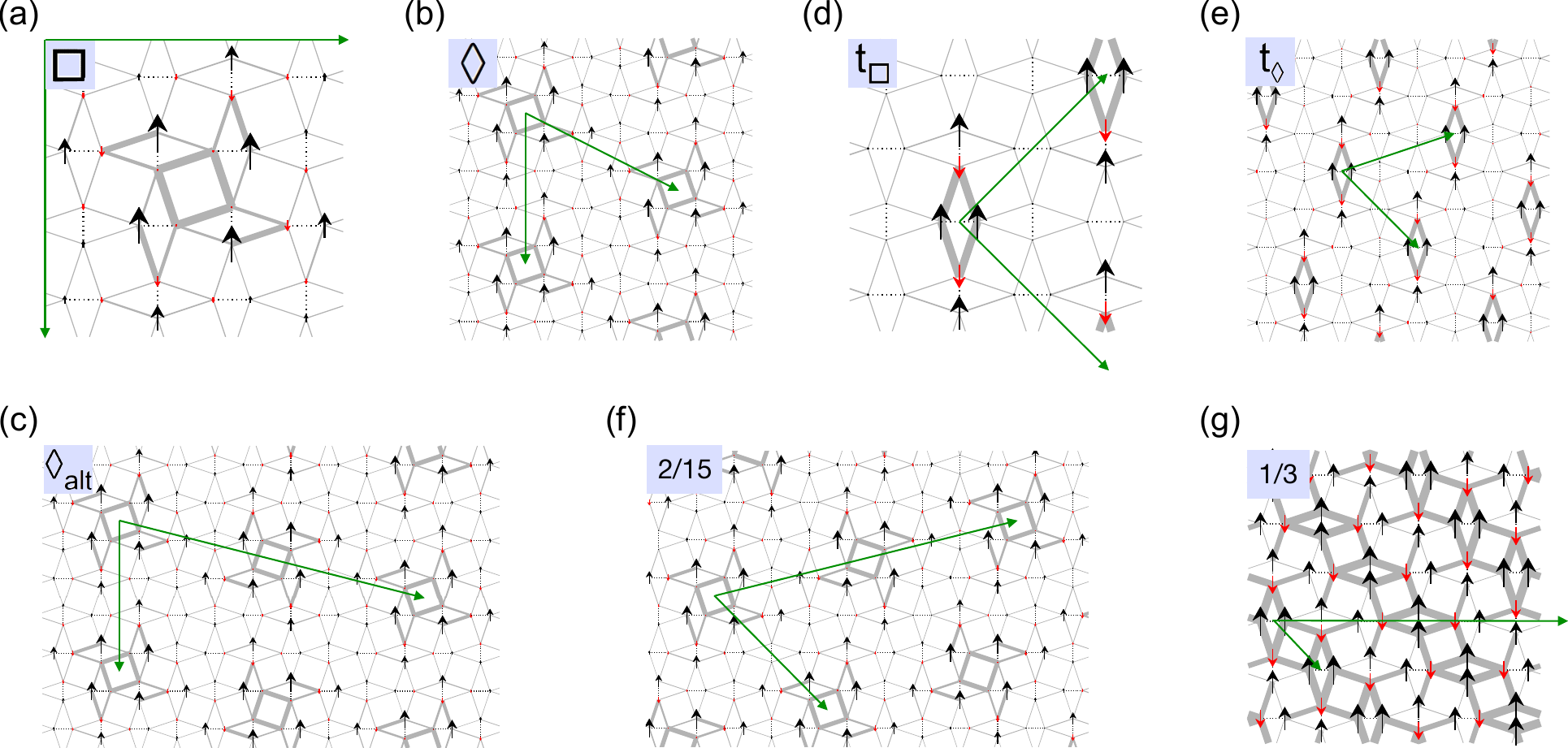}
    \caption{Spin structures for the (a-e) 1/8, (f) 2/15, and (g) 1/3 plateaus obtained with different unit cells in iPEPS. The unit-cell vectors are indicated by the green arrows. The thickness of the gray inter-dimer $J'$ bonds scales with the local bond energy, where thicker bonds correspond to lower energy. Dotted lines denote the intra-dimer $J$ bonds. The Supplementary Materials includes detailed information about the spin polarizations~\cite{SM}.}
    \label{Fig_structures}
\end{figure*}

\section{Material and the model}
\scbo is a layered antiferromagnet with a tetragonal structure, first synthesized in 1991 \cite{Smith1991}. In each layer, the Cu$^{++}$ ions form an orthogonal dimer arrangement (Fig.~\ref{Fig_structures}), well described by a Shastry-Sutherland lattice  \cite{Shastry1981}. The ratio $J'/J$ of the inter-dimer $J'$ and intra-dimer $J\simeq$~85~K interactions 
is estimated to be 0.63~\cite{Miyahara2003,Nomura2023}, for which a gapped singlet ground state is expected. Neglecting Dzyaloshinsky-Moriya (DM) interactions, $D/J \sim$~0.03-0.04~\cite{Kodama2005}, and inter-layer interactions, we get the two-dimensional (2D) spin Hamiltonian
\begin{equation}
H=J\sum_{\langle i,j \rangle}\bm S_{i}\cdot \bm S_{j}+J'\sum_{\langle \langle i,j \rangle\rangle}\bm
S_{i}\cdot \bm S_{j}- g_{zz} \mu_B H \sum_{i}S_i^z,
\label{eq:ssm}
\end{equation}
where $\langle i,j \rangle$ and $\langle \langle i,j \rangle\rangle$ denote respectively the nearest and the next nearest neighbors (NNN). $g_{zz}$, $\mu_B$, and $H$ are, respectively, the relevant component of the $g$ tensor, the Bohr magneton, and the magnetic field applied along the $z$ axis. The elementary triplet excitations (triplons) above the dimer-singlet ground state are almost localized, characterized by a nearly flat triplon band~\cite{Miyahara1999,Kageyama2000}. Because of the repulsive interactions between triplons, Wigner crystals of $S^z=1$ triplons are naturally stabilized at a finite magnetic field, as realized in the 1/4, 1/3 and 1/2 plateaus. In contrast to an isolated triplet, a pair of triplets is mobile thanks to correlated hopping, and the associated gain in kinetic energy leads to stable two-triplet bound states despite the repulsion~\cite{Momoi2000}. While the spin-2 bound states were initially expected to be relevant only in the dilute limit, it was later shown that they can crystallize and account for the low-magnetization plateaus below 1/4~\cite{Corboz2014}. 
We remark that the $J'/J$ value in \scbo is quite close to 
a phase transition into a valence-bond-solid plaquette phase at 0.675
\cite{Koga2000,Corboz2013}, which can be reached by applying hydrostatic pressure \cite{Waki2007, Haravifard2014, Zayed2017, Sakurai2018, Guo2020, Jimenez2021,Shi2022}. Here, we are rather focused on the effects of high magnetic fields, beyond 26~T, that induce the aforementioned cascade of magnetization plateaus before reaching saturation at $\simeq$~140~T~\cite{Nomura2023}. In the following, we determine the detailed magnetic structure in the 1/8, 2/15, and 1/3 plateaus, numerically and experimentally.

\section{Numerical results}

Our numerical results are based on infinite projected entangled-pair states (iPEPS)~\cite{verstraete2004,jordan2008}, a variational tensor network approach to represent ground states directly in the thermodynamic limit, see Supplementary Materials (SM)~\cite{SM}. The competing ground states are obtained by using different unit-cell sizes within iPEPS and by targeting specific magnetization sectors (1/8, 2/15, and 1/3). Similar calculations were performed in Refs.~\cite{Matsuda2013,Corboz2014}, however, here we employ a more accurate optimization method~\cite{liao19}, which yields lower variational energies and a higher precision in the individual spin magnitudes~\cite{corboz16b}, see SM~\cite{SM} for further details.

The iPEPS results for the spin structures are summarized in Fig.~\ref{Fig_structures}, with different candidate states for the 1/8 plateau presented in panels (a-e). The first panel illustrates the characteristic pinwheel-like structure of a bound triplet pair with $S^z=2$, obtained with a repeating square $4\times4$ unit cell. The rhomboid structure in Fig.~\ref{Fig_structures}(b) has a variational energy only slightly lower than that of the square structure, with an energy difference per site of only $2 \times 10^{-6} J$. An alternative rhomboid structure, displayed in Fig.~\ref{Fig_structures}(c), lies energetically between the two former states. While the variational energies of these states are extremely close, they are  clearly separated by $\approx 0.002 J$ from the candidate states corresponding to crystals of extended triplets, shown in Fig.~\ref{Fig_structures}(d,e), in agreement with previous results~\cite{Corboz2014}.
Finally, the structures of the 2/15 and 1/3 plateaus are presented in Figs.~\ref{Fig_structures}(f) and (g), respectively.
The former consists of bound triplet pairs, whereas the latter corresponds to a dense crystal of extended triplets arranged in diagonal stripes~\cite{Onizuka00,Matsuda2013}.

\begin{figure}[tb]
	\centering
	\includegraphics[width=\columnwidth]{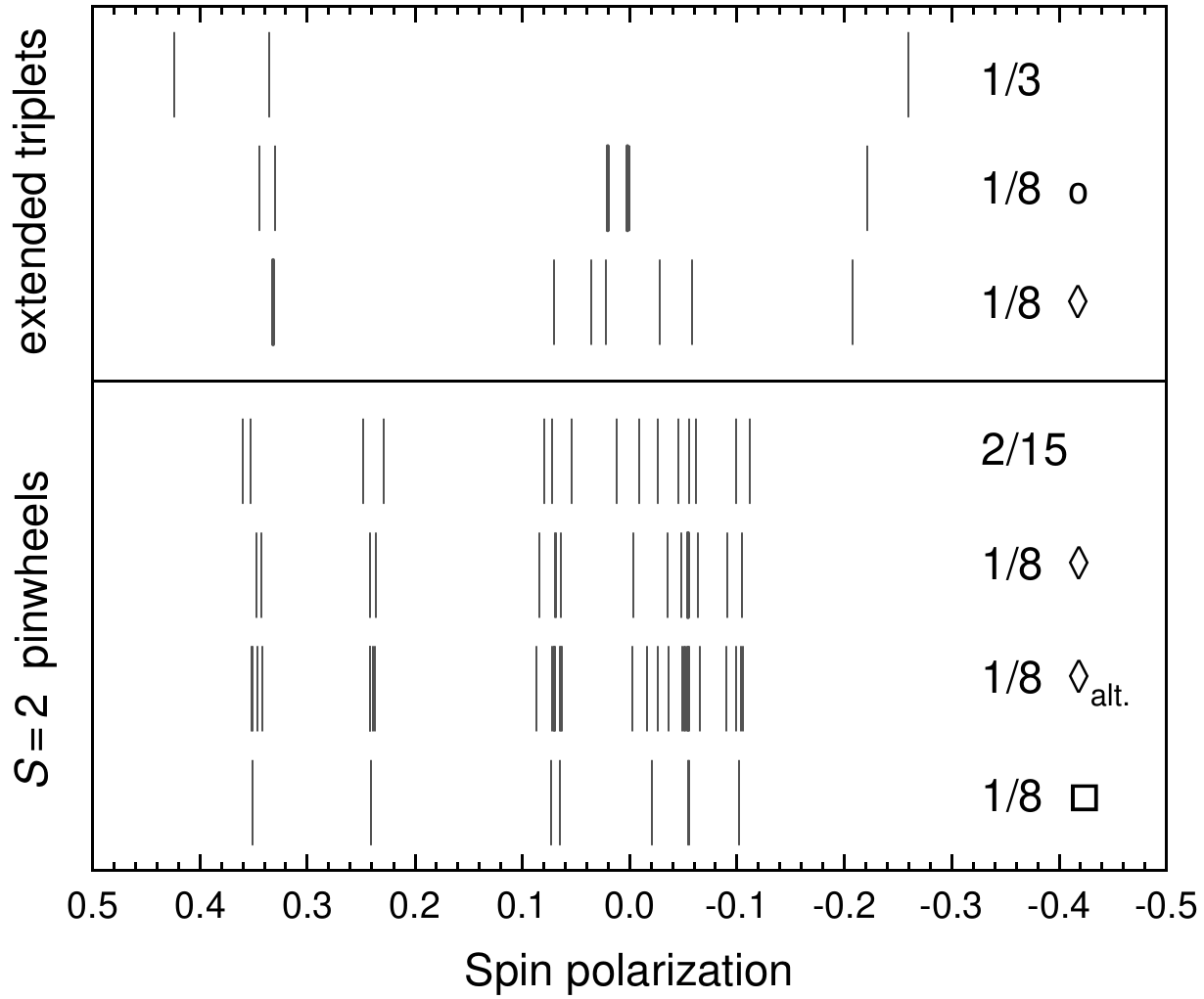}
	\caption{Histograms of the spin polarization values for the seven structures presented in Fig.~\ref{Fig_structures}: from top to bottom, the structures based on the extended triplets given in the panels (g), (d) and (e) and the structures based on the pinwheel-shaped spin-2 bound states given in the panels (f), (b), (c) and (a) of Fig.~\ref{Fig_structures}. Line thickness increases with the number of overlapping spin polarizations. Numerical values are provided in the SM~\cite{SM}.}
	\label{SpinPolarisations}
\end{figure}

In Fig.~\ref{SpinPolarisations} we present the corresponding histograms of the spin polarization values for the different structures, revealing strong qualitative differences between the extended triplet and pinwheel structures. The former exhibit strongly negatively polarized spins below -0.2 which are absent in the latter, where the lowest spin values are around -0.1. In contrast to the triplet structure of the 1/8 plateau, all pinwheel structures exhibit positively polarized spins in an intermediate range around 0.25, well below the most strongly polarized spins around 0.35. These characteristic features provide a clear distinction between the two classes of structures, and are directly reflected in the NMR spectra shown in Fig.~\ref{Spectra}. Even considering only the total span of the spin polarizations shown in Fig.~\ref{SpinPolarisations} compared to the total width of the spectra, spanning an interval of 470~MHz at 41.9~T and \emph{much} smaller intervals at lower fields, 330~MHz at 29.0~T and 310~MHz at 27.6~T, provides us with a strong indication that the two latter structures are based on pinwheels and not on extended triplets. To establish a definite proof that this is indeed the case, we proceed with the quantitative analysis of the NMR spectra in the following section. 

\section{NMR}

 As the magnetization plateaus in \scbo appear at very high magnetic fields, above 27 T, a full determination of their magnetic superstructures by structural techniques currently appears practically impossible. The only suitable technique being NMR spectroscopy, the first confirmation of the existence of a commensurate magnetic structure in the 1/8 plateau was provided by $^{63,65}$Cu NMR spectra \cite{Kodama2002a}. The advantage of using $^{63,65}$Cu nuclei is that they are on-site of the electronic Cu$^{++}$ spins and thus very strongly coupled to them. The local magnetic field values, $H_i$, determined from NMR spectra are thus dominantly proportional to the on-site local spin polarization $S_i$, and are only weakly capturing the polarization of the neighboring spins, whereas the dipolar coupling to the spins further away is negligible
\begin{equation}
\label{Couplings}
H_i/(g_{zz} \mu_B) = A\,S_i + C\,S_{\textrm{dimer}} + B \sum_{4~\textrm{NNN}} S_{\textrm{NNNi}},
\end{equation}
where $S_i$ refers to the on-site spin, $S_{\textrm{dimer}}$ to the other spin of the same dimer, $S_{\textrm{NNNi}}$ to the four NNN spins, and $|A|$, $|C|$, and $|B|$ are respectively the corresponding hyperfine fields~\cite{Kodama2002a}. For the given orientation of magnetic field, $H \parallel c$ axis, where $g_{cc} = 2.28$~\cite{Nojiri1999}, the expected $A$ value is $\approx$$-$20~T/$\mu_B$, while $C$ and $B$ are expected to be an order of magnitude smaller, $|A| \gg |C| > |B|$. The disadvantage of using $^{63,65}$Cu NMR spectra comes from the very high $A$ value, meaning that a fully polarized parallel or antiparallel spins $S^z = \pm 1/2$ would induce the local field values of about $\mp$23~T, making the observed NMR spectra very broad. Indeed, in the initial NMR investigation of the 1/8 plateau \cite{Kodama2002a}, the recorded $^{63,65}$Cu NMR signal spectrum spanned the frequency range of 100-420 MHz, and required an acquisition time that was prohibitively long for the time-constrained high-field measurements. All the subsequent NMR research was thus carried out using the $^{11}$B NMR \cite{Takigawa2008,Takigawa2010, Takigawa2013}, a nucleus having much weaker coupling to electronic spins and thus only $\approx$~9~MHz broad spectra of much stronger intensity. While these measurements allowed for clear identification of all the four first magnetization plateaus (1/8, 2/15, 1/6, and 1/4) \cite{Takigawa2013}, from $^{11}$B NMR spectra the determination of the corresponding magnetic structures is complex. The point is that much of the boron's coupling is of dipolar origin and thus extends quite far in space, capturing not only the in-plane contribution but also the contribution from neighboring planes. The boron spectra are thus more strongly affected by the unknown 3D stacking of the in-plane structure, leading to more free parameters.

\begin{figure*}[t!]
    \centering
    \includegraphics[width=0.80\textwidth]{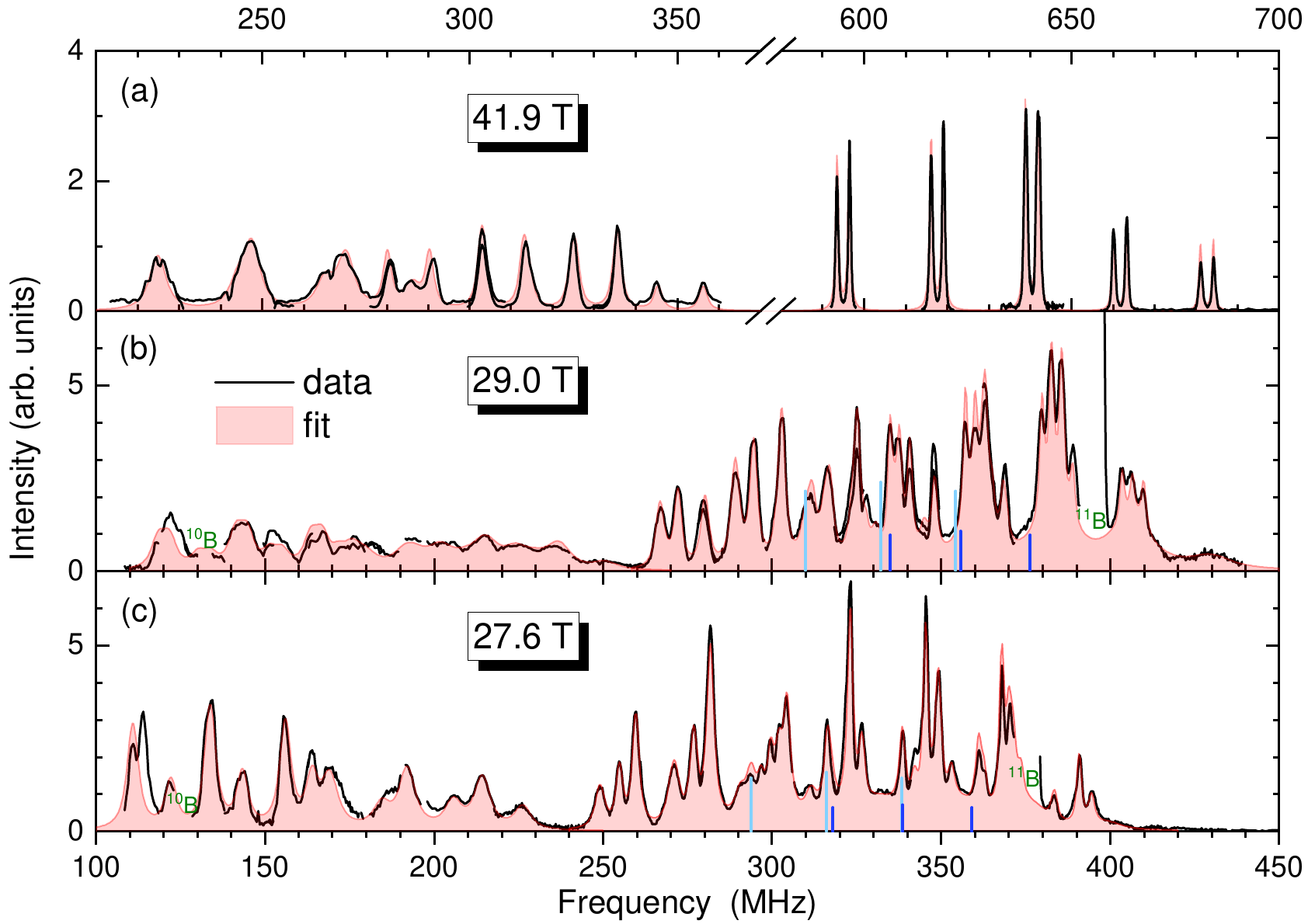}
    \caption{
    $^{63,65}$Cu NMR spectra (black solid lines) recorded at 0.5~K in the applied magnetic field of (a) 41.9~T and at 0.05~K in the field of (b) 29.0~T and (c) 27.6~T K, corresponding respectively to the 1/3, 2/15 and 1/8 magnetization plateau.
    Red line/surface is the fit explained in the text and (b,c) include sextuplets with $H_i=$~0~T, with light (dark) blue lines corresponding to $^{63(65)}$Cu. The $^{10}$B and $^{11}$B labels denote the positions where the boron NMR signal is largely dominant (not shown).
    }
    \label{Spectra}
\end{figure*}

\begin{figure}[t!]
	\centering
	\includegraphics[width=\columnwidth]{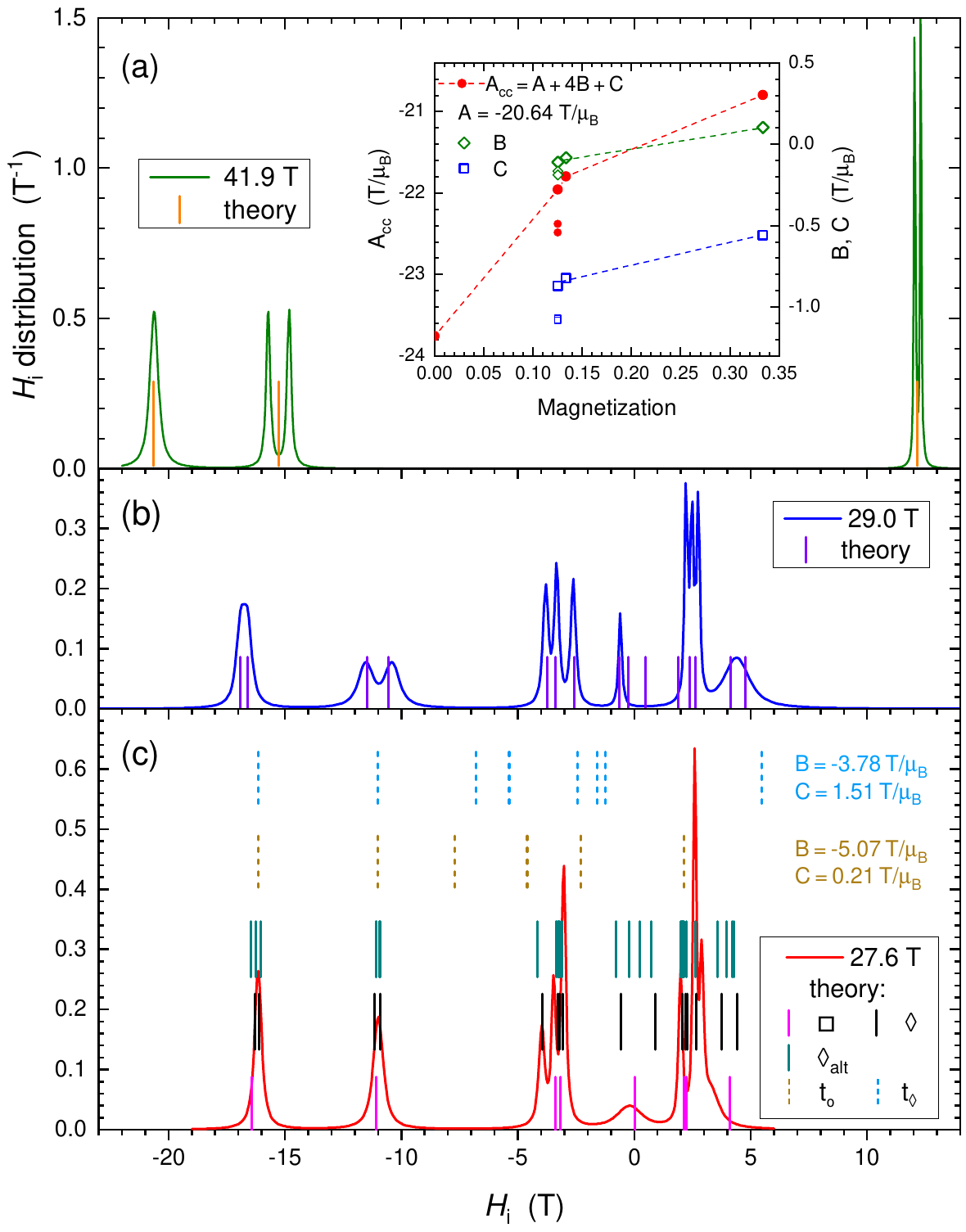}
	\caption{
Distribution of local fields obtained from the three $^{63,65}$Cu NMR spectra shown in Fig.~\ref{Spectra} (solid lines) are compared to values obtained from numerical simulations (vertical bars) for the 1/3 (a), 2/15 (b), and 1/8 (c) plateau.  Except for the 1/3 plateau, we obtained the best agreement by two-parameter fits that optimize the hyperfine fields $B$ and $C$, see the text for details. For the 1/8 plateau, five different (color coded) predictions are shown: three based on pinwheels making a square ($\Box$) or rhomboidal ({\large $\diamond$}, {\large $\diamond$}$_{\textrm{alt}}$) unit cells and two based on extended triplets (t$_{\Box}$, t{\large $_{\diamond}$}). The latter two fits, denoted by dashed lines, are clearly unsatisfactory and lead to nonphysical big $B$ values, signaling that these structures are inapplicable. The inset in panel (a) shows thus obtained hyperfine couplings. For the 1/8 plateau, smaller symbols correspond to the $\Box$ and {\large $\diamond$}$_{\textrm{alt}}$ structures presenting fits that are somewhat inferior than the optimal fit to the {\large $\diamond$} structure.
}
	\label{LocalFields}
\end{figure}

Fig.~\ref{Spectra} shows complete $^{63,65}$Cu NMR spectra recorded at 27.6 and 29.0~T in the M9 magnet of LNCMI-Grenoble using a top-loading dilution refrigerator (DR) and at 41.9~T in the LNCMI hybrid magnet using a $^3$He cryostat~\cite{Pugnat2026}. The employed temperatures, 0.05~K in DR and 0.5~K in the $^3$He cryostat, are low-enough to be considered as the zero limit for the magnetic structures of the corresponding 1/8, 2/15 and 1/3 plateaus, shown in Fig.~\ref{Fig_structures}.
The sample was placed in the $H \parallel c$ axis configuration, in a fixed orientation inside DR, and in a sample tilter inside the $^3$He cryostat. In the former case the sample tilt from the nominal orientation was estimated to be 1.2-1.6$^\circ$, from the splitting of the NMR line at 112~MHz in the spectrum shown in Fig.~\ref{Spectra}c \cite{SM}, while in the latter case the tilt was reduced to below 0.5$^\circ$. The employed sample mountings enabled us to minimize the distance between the NMR excitation coil with the sample and the semi-rigid coaxial cable, and thus improve the high-frequency signal. The signal was further improved by using better top-tuning and decoupler circuitry to ensure full 50 $\Omega$ continuity. 
The spectra were recorded as overlapping, ~10\% broad, automatic frequency sweeps, described in Ref. \cite{Li2026}. Altogether, as compared to the previously published $^{63,65}$Cu NMR spectra at 27.6 T \cite{Kodama2002a}, on the same sample, the newly recorded spectra present a \emph{much} higher signal-to-noise ratio and precision, included at the high-frequency end. This enabled a more reliable determination of the local field values as follows:

As the two $^{63,65}$Cu isotopes are spin-3/2 nuclei, for each Cu site, each isotope presents three NMR lines whose frequencies $f$ are equidistant for the given orientation of magnetic field, split by the quadrupolar coupling $^{63,65}\nu_c$ \cite{Kodama2002a,Abragam1961}:
\begin{equation}
\label{Triplet}
f(\alpha,m) =\,^{\alpha}\gamma\,[(1+K_c)H + H_i] + m\,^{\alpha} \nu_c,
\end{equation}
where $\alpha$~= 63 or 65 denotes the isotopes, $m$~= 0 for the central line and $\mp$1 for the two ``satellites'', $^{\alpha}\gamma$ is gyromagnetic ratio of 11.285 (12.089) MHz/T, and $K_c$~= 1.69\%, the known orbital shift of the lines \cite{Kodama2002b}. The intensity of the central line is expected to be $\sqrt{4/3}$~= 1.155 stronger than that of the satellites, and the relative intensities for the two isotopes are defined by their relative abundance $^{63}n/^{65}n$~= 0.691/0.309~= 2.24. For each local field value $H_i$, we thus have a sextuplet of NMR lines whose position is defined only by the $H_i$, as all the other parameters and relative intensities are constant and known, see the $H_i = 0$ sextuplet examples in Fig.~\ref{Spectra}(b,c).
To reproduce the observed shapes of each line, we used the Voigt line shape, where the ratio of Lorentzian and Gaussian widths was fixed to 2.0. Each sextuplet of lines is thus fully defined by only two parameters, the $H_i$ value and the corresponding Gaussian width that measures the spread of this local value. The relative intensity of the sextuplets was taken to be integer (1 or 2), with the sole exception of the $H_i \approx$~0 line at 29~T, where it was 0.5, as the line intensity was apparently reduced by shorter $T_2$ relaxation. Fitting a spectrum with thus defined sextuplets amounts to a ``deconvolution'' that reduces the complicated spectrum with too many overlapping lines into a much simpler, sought-for information on the distribution of the local fields only, shown in Fig.~\ref{LocalFields}. As a matter of fact, this distribution is the NMR spectrum we would observe if copper nuclei had only one isotope of spin-1/2, insensitive to quadrupolar coupling.

The predicted magnetic structure of the 1/3 plateau is very simple [Fig.~\ref{Fig_structures}(g)], presenting ``stripes'' of only three different spin polarizations values, which can be regarded as the close packing of extended triplets. The important point is that this structure is not controversial: all numerical approaches predict the same structure, with three distinct spin polarization values~\cite{Onizuka00,Miyahara2003,Matsuda2013,Corboz2014}.
To these three values should correspond a quite simple NMR spectrum presenting only three corresponding $H_i$ values. 
At 41.9~T, the experimental data indeed show three different $H_i$ values [Figs.~\ref{Spectra}(a) and \ref{LocalFields}(a)], whereas the middle and the high-frequency ones are somewhat split in two, to $\mp$3.1\% and $\mp$1.0\% of their respective average values. These splittings should be related to the 3D stacking of the spin planes and, as such, they are not covered by the simulations. In the SM~\cite{SM} we show that this configuration of the line splittings corresponds to an ABAB type of 3D stacking. Neglecting the 3D effects, we used average position of the split $H_i$ values for the analysis. First, using Eq.~(\ref{Couplings}) it is possible to prove that the ratio of the average spin polarizations and the average local fields defines the average hyperfine field, $A_{cc}(q$\,=\,$0) = A + C + 4B = \langle H_i \rangle / (g_{cc} \mu_B \langle S_i \rangle)$. We thus determine the $A_{cc}(q$\,=\,$0)$~= $-$20.80~T/$\mu_B$ value. Respecting this constraint, we then solve for the three experimental $H_i$ values using theoretical spin polarization values and the structure, leading to $A, C, B =  -20.64, -0.56, 0.10$~T/$\mu_B$ (Fig.~\ref{LocalFields}(a)). The obtained on-site value $A$ is nearly the same as in other compounds for a Cu$^{++}$ electronic spin in the $d_{x^2-y^2}$ orbital surrounded by four oxygen ions, e.g., in CuGeO$_3$, $A_{zz}$~= $-$19.9~T/$\mu_B$ \cite{Fagot-Revurat1997}. As expected, the transferred $C$ and $B$ terms are much smaller and strongly decrease with the distance: $|C/A|$ = 2.7\% and $|B/A|$ = 0.5\%, confirming that the coupling is indeed dominantly on-site with only a small correction from neighboring sites. We remark that the $A_{cc}(q$\,=\,$0)$ value we determined at 41.9~T has 12\% smaller magnitude than the one determined at a low field of 8~T from the shift versus magnetic susceptibility plots ($-$23.76~T/$\mu_B$)~\cite{Kodama2002b}. The difference is attributed to the magnetic field dependence of the transferred $C$ and $B$ terms, corresponding to magnetostriction, while the local on-site $A$ is expected to be field independent.  

This $A$ value is then used in fitting the experimental $H_i < -1.5$T values found in the 2/15 and 1/8 plateaus [Figs.~\ref{LocalFields}(b) and (c)], respectively at 29.0 and  27.6~T, with the parameters of the least-squares fit being only $C$ and $B$. For all pinwheel-based structures, which were not known during the analysis of Ref.~\cite{Kodama2002a}, this leads to remarkably precise fits of all $H_i < 0$ values (i.e., positive local spin values) and very good coverage of $H_i > 0$ values. Regarding these latter values, we note that 27.6~T spectra are ``missing'' lines predicted at $\approx$~0~T and $\approx$~4~T, while the 29.0~T spectrum is missing 2.5 line intensity at $\approx$~0~T. This wipe-out is probably related to faster fluctuations of the corresponding spins (in particular the nearly zero-polarized ones), inducing shorter $T_2$ relaxation. Incidentally, these missing lines in the spectra prevent us from determining the $A_{cc}$ values, which would otherwise further constrain the fits. We observe that the effects of the 3D coupling that are clearly visible in the dense 1/3-plateau structure are here apparently absent, probably because these spin structures are quite diluted. Furthermore, the effects of the neglected DM terms \cite{Kodama2005} are also apparently very small. Finally, comparing the quality of the fits for the three different structures of the 1/8 plateau, we note that the best fit is provided for the rhomboid structure, which is also theoretically the most stable one. However, taking into account the employed approximations (no 3D effects nor DM couplings), it is not clear whether this selection is really significant.

For the 1/8 plateau, we also fitted the predicted extended-triplet-based structures (dashed vertical bars in Fig.~\ref{LocalFields}(c)), where the two fit parameters were used to ensure a perfect fit of the two most negative $H_i$ values. Although this is formally always possible, compared to the values of the 1/3 plateau, the fitted $|B|$ values are much too high and unexpectedly higher than the $|C|$ values, altogether pointing to a nonphysical fit. This is further corroborated by the globally poorer fit for all the other predicted $H_i$ values, where the predicted \emph{isolated} most positive $H_i$ value, clearly observed at 41.9~T, is totally absent in the measured lower field spectra. Altogether, our data clearly reject the extended-triplet-based structures in the 1/8 plateau.

\section{Discussion}
The $S^z = 1$ extended triplets are very compact, mainly de-localized over only 3 neighboring dimers [Fig.~\ref{Fig_structures}(d)], and can be easily understood within a classical picture: When an up-up configuration of spin polarization is induced on a dimer, this breaks the frustration in the interaction to the two neighboring orthogonal dimers and drives them into a down-up configuration. These AF ``wings” thus carry only small net polarization, but they notably lower the energy of the extended triplet, while keeping it isolated from other spins by frustration. They are thus prone to be stable in the rather densely packed structures of the moderately-magnetized 1/4 and 1/3 plateaus, which is confirmed by both the theoretical simulations~\cite{Miyahara2003,Corboz2014} and our 1/3-plateau NMR spectrum.

In contrast, the spin-2 pinwheels are mainly de-localized over as many as 8 dimers [Fig.~\ref{Fig_structures}(a)], they are manifestly of quantum nature and cannot be understood in classical terms. 
Energetically, they are favored especially for large $J'/J$~\cite{Corboz2014} thanks to the formation of low-energy bonds on the center plaquette, reminiscent of the plaquette phase stabilized at zero magnetic field for $0.675(2) \le J'/J \le 0.785(5)$~\cite{Koga2000,Corboz2013,Corboz2026}, and by avoiding pairs of parallel spins on dimers, which in triplet crystals come with a high energy cost.
Low-magnetization 1/8, 2/15 and 1/6 plateaus allow for enough space that these objects require, but it is a priori not clear whether in a real compound they remain energetically favorable despite neglected Hamiltonian terms, such as DM-terms and 3D couplings. Through comparison with the numerical spin polarizations, our NMR spectra in the 1/8 and 2/15 plateaus demonstrate that this is indeed the case and thus firmly establish the existence of such complex spin-2 objects. To our knowledge, this is the only case where spin-2 (triplon pair) excitations are directly observed in real space, proving the unique capabilities of NMR in an extreme environment of high magnetic fields. 

The present results demonstrate the experimental relevance of a new paradigm where magnetization plateaus are neither the result of the stabilization by quantum fluctuations of a classical collinear state with up and down spins in a finite field range, nor the localization of the \emph{elementary} quasiparticles of the systems (triplons in dimer-based systems or magnons below saturation), but are the consequence of a two-step process where a more complex object forms (here a spin-2 bound state) and is localized by interactions. Work is in progress to try and identify other instances of this mechanism in dimer-based antiferromagnets, such as the maple-leaf model~\cite{Ghosh2022,Ghosh2023}.

The experimental proof that spin-2 bound states crystallize in \scbo demonstrates the subtlety of the interplay between kinetic energy and repulsive interactions in flat band quantum systems. The gain in kinetic energy of pairs of particles is responsible for the formation of bound states in the first place because single particles cannot move.
Yet, these bound states can give up some of their kinetic energy to minimize long-range interactions without unbinding~\cite{Corboz2014,Fogh2024}. This mechanism opens new perspectives on the formation of magnetization plateaus and, more generally, on the phase diagrams of flat-band systems beyond frustrated magnetism.

\begin{acknowledgments}
The authors would like to acknowledge valuable discussions with Raivo Stern and Hannes K\"uhne. We thank the hybrid magnet team at LNCMI and CEA that made the 41.9~T measurements possible: P. Pugnat, R. Barbier, F. Debray, C. Grandcl\'ement, Y. Krupko, F. Molini\'e, K. Paillot, R. Pankow, R. Pfister, L. Ronayette, B. Vincent and Ch. Simon.
This work was supported by LNCMI-CNRS, a member of the European Magnetic Field Laboratory (EMFL), and by the Swiss National Science Foundation under grant number 212082. 
This project has received funding from the European Research
Council (ERC) under the European Union’s Horizon
2020 research and innovation programme (Grant Agreement
No. 101001604).
\end{acknowledgments}

\end{document}


\title{Supplementary Materials for\\ 
Pinwheel-shaped bound triplet pairs in the magnetization plateaus of SrCu$_2$(BO$_3$)$_2$}

\author{Igor Vinograd\,\orcidlink{0000-0002-1054-3741}}
\thanks{These authors contributed equally to this work.}
\affiliation{
Laboratoire National des Champs Magn\'{e}tiques Intenses, LNCMI-CNRS (UPR3228), EMFL, \\
Univ. Grenoble Alpes, Univ. Toulouse, INSA-T, 38042 Grenoble Cedex 9, France}

\author{Philippe Corboz\,\orcidlink{0000-0003-1024-1230}}
\altaffiliation{These authors contributed equally to this work.}
\affiliation{
Institute for Theoretical Physics,
University of Amsterdam, Science Park 904, 1098 XH Amsterdam, The Netherlands}

\author{Steffen Kr\"{a}mer\,\orcidlink{0000-0002-6107-3583}}
\affiliation{
Laboratoire National des Champs Magn\'{e}tiques Intenses, LNCMI-CNRS (UPR3228), EMFL, \\
Univ. Grenoble Alpes, Univ. Toulouse, INSA-T, 38042 Grenoble Cedex 9, France}

\author{Yanan Li\,\orcidlink{0000-0003-3851-9425}}
\affiliation{
Laboratoire National des Champs Magn\'{e}tiques Intenses, LNCMI-CNRS (UPR3228), EMFL, \\
Univ. Grenoble Alpes, Univ. Toulouse, INSA-T, 38042 Grenoble Cedex 9, France}
\affiliation{
Present address: College of Electronic and Information Engineering, Liaoning Technical University,
125105, Huludao, Liaoning, China}

\author{Fr\'ed\'eric Mila\,\orcidlink{0000-0003-4306-7996}}
\affiliation{
Institute of Physics, \'{E}cole Polytechnique F\'{e}d\'{e}rale de Lausanne (EPFL), CH-1015 Lausanne, Switzerland}

\author{Hiroshi Kageyama\,\orcidlink{0000-0002-3911-9864}}
\affiliation{
Graduate School of Engineering, Kyoto University, Nishikyo-ku, Kyoto 615-8510, Japan}

\author{Masashi Takigawa\,\orcidlink{0000-0002-4516-5074}}
\affiliation{
Professor Emeritus, Institute for Solid State Physics, University of Tokyo, 5-1-5 Kashiwanoha, Kashiwa, Chiba 277-8581, Japan}

\author{Mladen Horvati{\'c}\,\orcidlink{0000-0001-7161-0488}}
\email{mladen.horvatic@lncmi.cnrs.fr}
\affiliation{
Laboratoire National des Champs Magn\'{e}tiques Intenses, LNCMI-CNRS (UPR3228), EMFL, \\
Univ. Grenoble Alpes, Univ. Toulouse, INSA-T, 38042 Grenoble Cedex 9, France}
\date{\today}

\maketitle


\section{Methods}
An infinite projected entangled-pair state (iPEPS) is a variational tensor network ansatz to represent 2D ground states in the thermodynamic limit~\cite{verstraete2004,jordan2008}, with an accuracy that is systematically controlled by the bond dimension $D$ of the tensors. The ansatz consists of a unit cell of tensors that is periodically repeated on a square lattice, here with one tensor per dimer. By using different unit cell sizes, different low-energy states can be obtained, and the ground state can be identified from the state with the lowest variational energy. The contraction of the two-dimensional tensor network is done based on a variant~\cite{corboz2011,corboz14_tJ} of the corner transfer matrix renormalization group method~\cite{nishino1996,orus2009-1}, where we exploit the U(1) spin symmetry of the model to increase efficiency~\cite{singh2010,bauer2011} and to target specific magnetization sectors. To determine the optimal parameters, we employ a combination of the simple-update~\cite{jiang2008} and full-update imaginary time evolution approaches~\cite{jordan2008,phien15},  further refined by a direct energy minimization based on automatic differentiation~\cite{liao19} for bond dimensions up to $D=8$ for the 1/8 plateau and $D=6$ for the 2/15 and 1/3 plateaus. A similar approach applied to the SSM at zero magnetic field has yielded the lowest variational energies to date~\cite{Corboz2026}.


\section{Additional iPEPS results}
\begin{figure}[h!]
    \centering
    \includegraphics[width=0.9\textwidth]{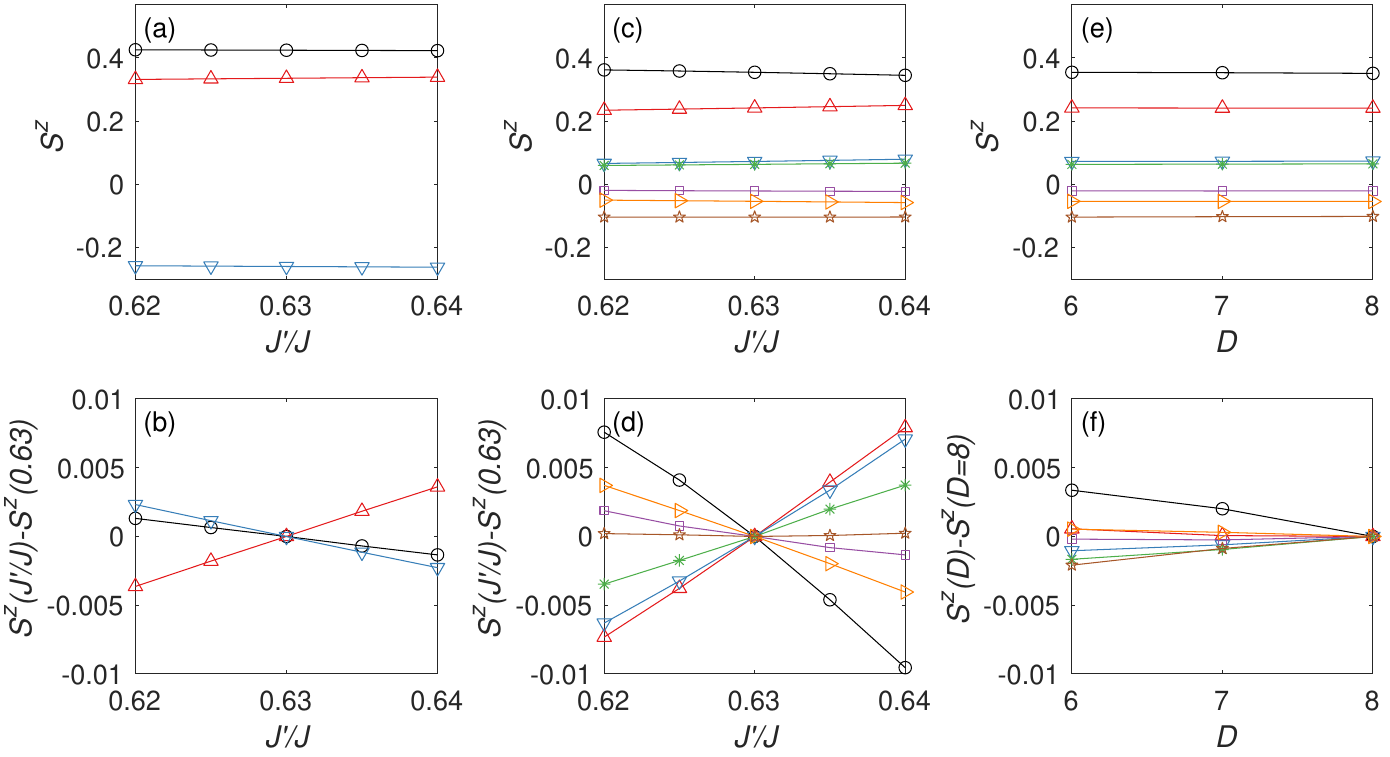}
    \caption{(a) Local spin values of the 1/3 magnetization plateau as a function of $J'/J$ ($D=6$). (b) Differences in the local spin values with respect to the reference values at $J'/J=0.63$ for the 1/3 plateau.  (c,d) Same as  (a,b), but for the 1/8 magnetization plateau (square structure), for $D=8$. (e) Same as (c), but plotted as a function of the bond dimension $D$ at $J'/J=0.63$. (f) Same as (d), but relative to the values at the largest bond dimension $D$ at $J'/J=0.63$.
    }
    \label{fig_iPEPSresults}
\end{figure}

In the main text, a coupling ratio of $J'/J=0.63$ was used, which in Ref.~\cite{matsuda13} was obtained from a fit to the experimental magnetization curve at high magnetization values $m\ge 1/4$, with an estimated uncertainty of approximately $\Delta\sim0.005$. In Fig.~\ref{fig_iPEPSresults}(a), we present results illustrating the sensitivity of the spin values to the coupling ratio in the range $J'/J\in[0.62,0.64]$ for the 1/3 magnetization plateau, revealing only a very weak dependence. In Fig.~\ref{fig_iPEPSresults}(b) we show the same data with the reference values at  $J'/J=0.63$ subtracted. Figures~\ref{fig_iPEPSresults}(c) and (d) display similar results for the spin values of the 1/8 plateau (square structure), showing a slightly stronger dependence, but still well below 1\%. 
For comparison, we show  in Fig.~\ref{fig_iPEPSresults}(e) and (f) the  spin values as a function of the bond dimension $D$ for the 1/8 plateau (square structure),  which for a gapped state are expected to converge exponentially with $D$. Finite bond dimension effects are found to  be very small at the large bond dimensions considered here. 

\section{Spin polarization values}
\begin{figure}[h!]
    \centering
    \includegraphics[width=0.88\textwidth]{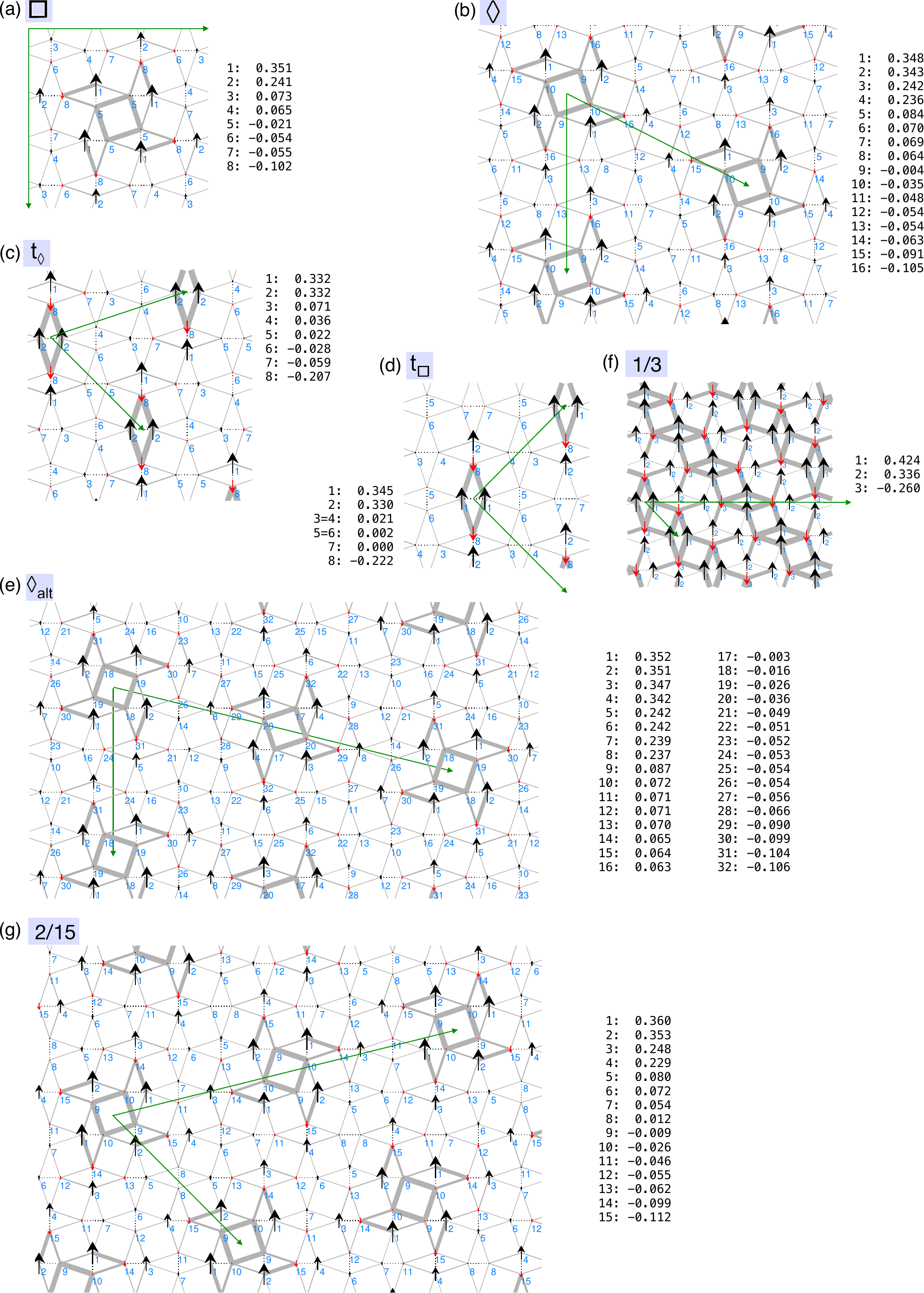}
    \caption{Spin polarization values of all spin structures shown in Fig.~2 in the main text, at all non-equivalent sites and sorted in descending order.
    }
    \label{fig_spinvalues}
\end{figure}

In Fig.~\ref{fig_spinvalues} we present the spin polarization values for each spin structure. The values are averaged over the symmetry-equivalent sites within each structure. Histograms of these spin polarizations are shown in Fig.~2 in the main text.

\section{Determination of the sample tilt in the dilution refrigerator}

For the experiments in the 1/8 and 2/15 plateaus, recorded at 27.6~T and 29.0~T in a dilution refrigerator, the sample was placed in the probe in a fixed orientation, which typically results in a misorientation of up to $2-$$3^{\circ}$. Such a tilt should in principle induce a line splitting of the NMR spectra, which will be visible only if it is bigger than the width of the lines. In the $^{63,65}$Cu spectrum at 27.6~T shown in Fig.~3 of the main text, we notice a visible line splitting only for the lowest frequency line, which points to some specific, magnetic-field and nuclear-transition dependent mechanism. Indeed, for this line, the spin-induced field is strongly negative, leading to a total local field comparable to the quadrupolar frequency. The quadrupolar line splitting is then no longer precisely equidistant and becomes increasingly dependent on the angle. In our case, the angle dependence is dominantly generated as the very strong anisotropy of the hyperfine coupling tensor $A_{\perp}/A_{\|} \simeq 0.1$ acts on a big $H_i$ value, thus producing the angle dependent total local field. In this case, to precisely fit the line positions of each $^{63,65}$Cu sextuplet of lines that corresponds to a certain $H_i$ value, one has to numerically solve a complete, Zeeman plus quadrupolar, spin Hamiltonian. In this way we find that the observed splitting of 2.9 MHz of the lowest frequency line corresponds to a tilt of 1.6$\mp$0.4$^{\circ}$ for the anticipated $A_{\|}/A_{\perp}$ values of 0.1$\mp_{0.1}^{0.2}$. For these tilt values, the line splitting for other nuclear transitions is much smaller, and also strongly decreases when the $H_i$ value is decreased. This makes that the splittings of all other lines in the 27.6~T spectrum, and all the lines in the 29.0~T spectrum, are ``covered'' by the width of the lines and thus unobservable.           

\section{Internal field values for the simulation of $^{63,65}$Cu NMR spectra}
Each local field $H_i$ value corresponds to a sextuplet of $^{63,65}$Cu lines whose positions are defined by Eq.~(3) of the main text, whereas their line shape was taken to be the Voigt function having the Lorentzian to the Gaussian width ratio fixed to $w_L/w_G = 2.0$.
The spectra were simulated by a sum of the sextuplets having integer values of the relative amplitudes $A$ and the following local fields values parameters. All parameters are summarized in Table~\ref{tab:63Cu_internal_field}.
To determine the hyperfine coupling by fitting the calculated spin polarizations to the measured local fields $H_i < -1.5$~T, for some magnetic structures weakly separated calculated spin polarizations are grouped together and averaged such that the number of averaged calculated polarizations is equal to the number of experimentally observed lines. Note that for the 1/8 plateau the best fit is obtained for the rhombohedral structure with 16 sites. Out of the two lines predicted at about -1 T and +1 T, we observe only one broad line near 0 T (line N°6). The line predicted at about +4 T is not observed. Altogether, this means there are two lines whose intensity is strongly reduced by short $T_2$.
For the 2/15 plateau at 29 T there are three lines predicted for $|H_i| <$ 1 T, and only 0.5 of the lines are observed, with the remaining 2.5 “missing” lines again reduced by short $T_2$.

\begin{table}[htbp]
\centering
\small
\begin{tabular}{|c|ccc|ccc|ccc|}
\hline
 & \multicolumn{3}{c|}{\textbf{27.6 T}} & \multicolumn{3}{c|}{\textbf{29.0 T}} & \multicolumn{3}{c|}{\textbf{41.9 T}} \\
\cline{2-10}
\textbf{Line N°} & $A$ & $H_i$ / [T] & $^{63}w_G$ / [MHz] & $A$ & $H_i$ / [T] & $^{63}w_G$ / [MHz] & $A$ & $H_i$ / [T] & $^{63}w_G$ / [MHz] \\
\hline
1  & 2 & -16.15 & 1.7  & 1 & -16.9  & 2.0  & 1   & -19.282 & 2.0  \\
2  & 2 & -10.99 & 2.4  & 1 & -16.6  & 2.0  & 0.5 & -14.382 & 1.0  \\
3  & 1 & -3.97  & 1.4  & 1 & -11.53 & 3.5  & 0.5 & -13.482 & 1.0  \\
4  & 1 & -3.46  & 1.0  & 1 & -10.4  & 3.5  & 0.5 &  13.383 & 0.35 \\
5  & 2 & -3.03  & 1.05 & 1 & -3.8   & 1.3  & 0.5 &  13.643 & 0.33 \\
6  & 1 & -0.2   & 6    & 1 & -3.33  & 1.1  & $\Sigma_A = 3/3$ & & \\
7  & 1 &  1.99  & 0.9  & 1 & -2.62  & 1.2  &     &         &      \\
8  & 2 &  2.6   & 0.75 & 0.5 & -0.6 & 0.8  &     &         &      \\
9  & 1 &  2.9   & 0.9  & 1 &  2.23  & 0.7  &     &         &      \\
10 & 1 &  3.3   & 4    & 1 &  2.48  & 0.8  &     &         &      \\
11 & $\Sigma_A = 14/16$ & &  & 1 &  2.75  & 0.75 &     &         &      \\
12 &   &        &      & 2 &  4.4   & 6    &     &         &      \\
   &   &        &      & $\Sigma_A = 12.5/15$ & & &     &         &      \\
\hline
\end{tabular}
\caption{Internal field values, $H_i$, their widths, $^{63}w_G$, and amplitudes, $A$, used to simulate the $^{63,65}$Cu NMR spectra. The $\Sigma_A$ denotes the observed over the expected number of lines, whereas the missing lines are ``wiped-out'' by fast $T_2$ relaxation. Measurements in the 1/8 plateau at 27.6~T and in the 2/15 plateau at 29.0~T were made in a dilution refrigerator at 50 mK. The 1/3 plateau was measured at 41.9~T and 0.4 K.}
\label{tab:63Cu_internal_field}
\end{table}

\section{Determination of the 3D stacking in the 1/3 plateau}

\begin{figure}[h!]
    \centering
    \includegraphics[width=0.8\textwidth]{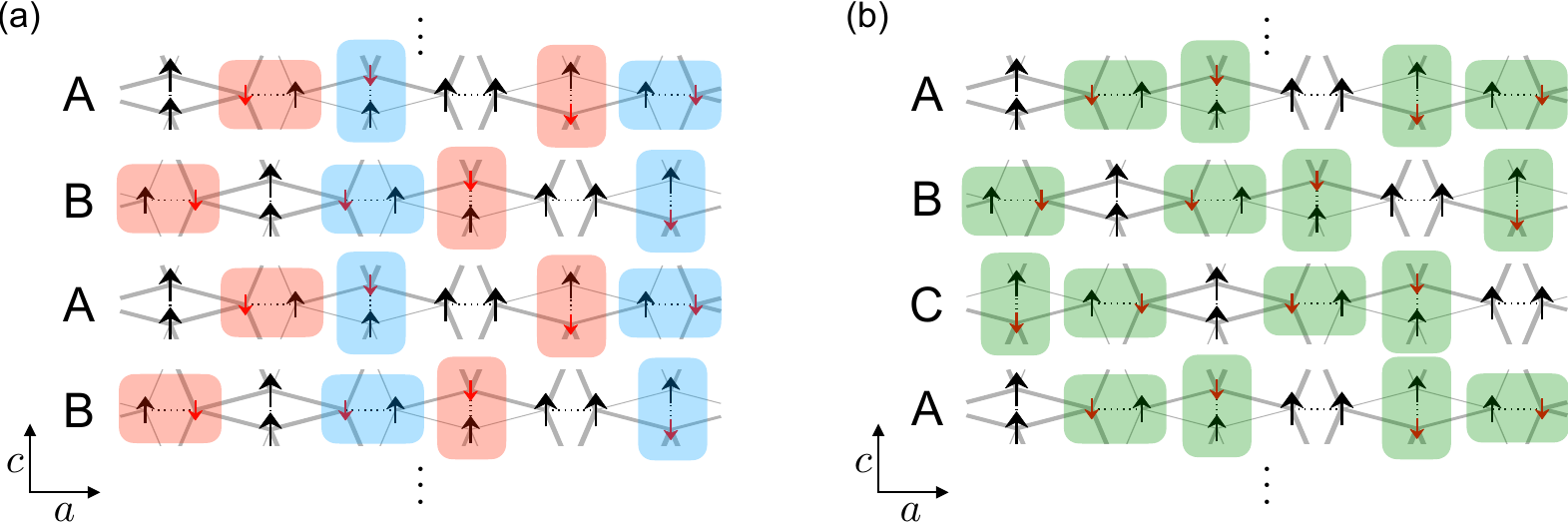}
    \caption{ (a) Visualization of ABAB stacking of four Cu layers vertically. Strongly polarized triplets always alternate with wings of extended triplets from adjacent planes (red dimers). Neighboring wings (blue dimers) are stacked on top of each other, leading to a different local environment for Cu spins belonging to red and blue wings. (b) ABC stacking leads to identical local environments for all dimers. Strongly polarized triplets are always adjacent to wings (green dimers) and wings are always adjacent to a triplet and another wing, so no splittings are expected for ABC stacking.}
    \label{stacking}
\end{figure}

Theoretically, only three local fields are expected in the $^{63,65}$Cu spectra of the 1/3 plateau, but we observe that two of them are slightly split, leading to 5 different sites, related to inter-layer stacking effects. From this we deduce ABAB stacking as the only one in which the wings of the extended triplets of a layer A lie between the two triplets or in between the two wings of the triplets of the two adjacent layers B. This case is shown in Fig.~\ref{stacking}(a). Note that strongly polarized triplets always lie adjacent to wings, so the corresponding local field $H_i$ of a strongly polarized Cu spin is not split. Contrary to that, the ABC stacking makes that all the wings of the triplets have both wings and triplets in the two adjacent planes. This makes the local environment of each of the three Cu spins equivalent. No splitting of local fields would be observed in this case, and is thus inconsistent with the experimental data.

\section{$^{11}$B-NMR in the 1/3 plateau}

\begin{figure}[b]
    \centering
    \includegraphics[width=0.9\textwidth]{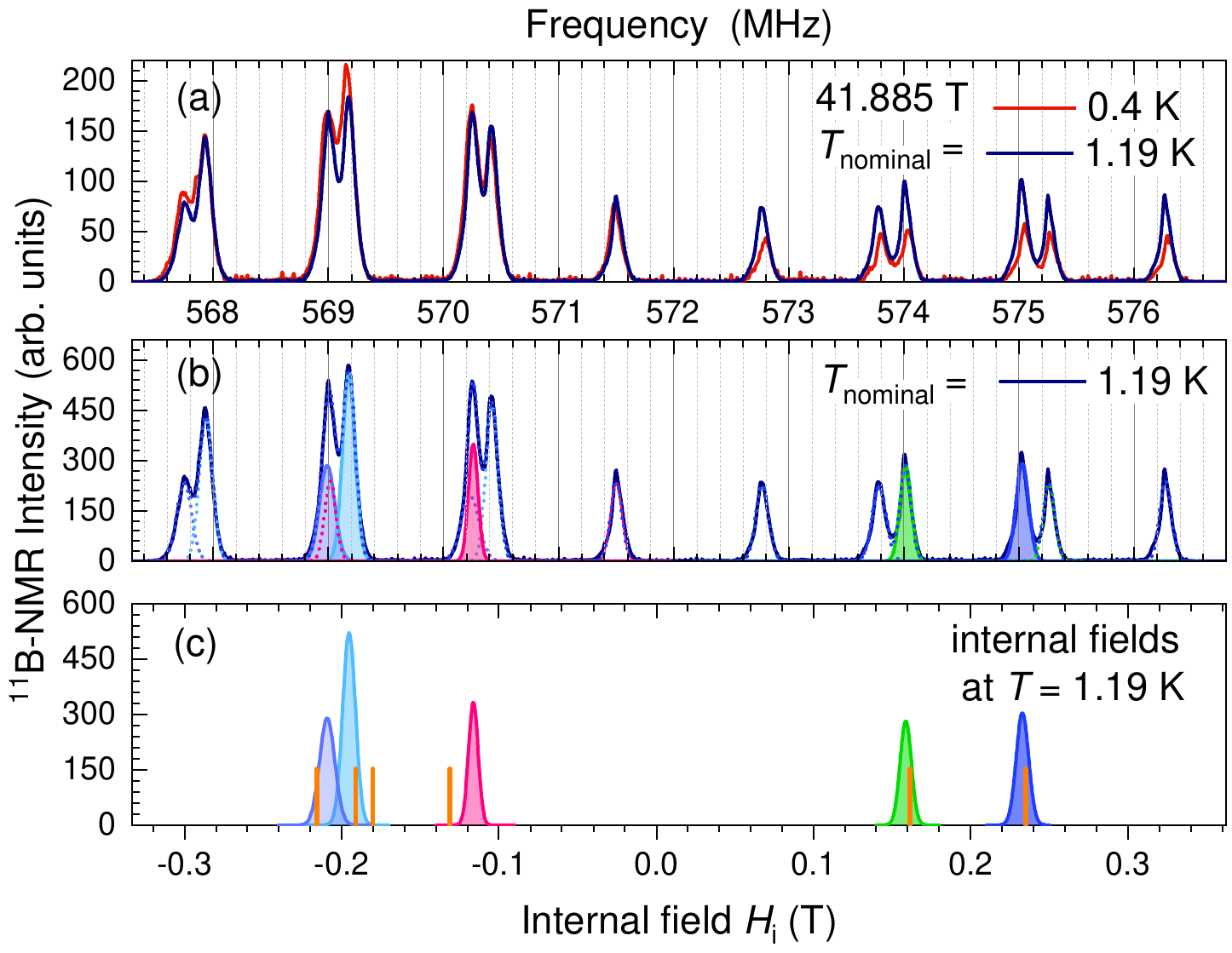}
    \caption{(a) $^{11}$B-NMR in the 1/3 plateau measured as frequency sweeps at $T=$0.4 K and $T=$1.19 K. (b) Fits of the the spectrum at $T=$1.19 K which allow to separate quadrupolar satellites (dotted lines) and to identify the central peaks (filled colours). (c) Internal fields, $H_i$, are inferred from the central peaks from (b) using the gyromagnetic ratio $^{11}\gamma$=13.66 MHz/T and are corrected for a demagnetizing field of $H_d=-0.022$~T. Orange bars are fits to the peak positions. Relative intensities at 1.19 K and 0.4 K might vary due to peak dependent $T_1$ and $T_2$.
    }
    \label{11B_NMR}
\end{figure}

We have measured $^{11}$B NMR in the 1/3 plateau, which can be compared to previous data. $^{11}$B carries nuclear spin $I=3/2$ which similarly to Cu NMR leads to three transitions, giving a central peak and two satellites for each local environment.
In Fig.~\ref{11B_NMR}(a) we show that the spectral features change only slightly between 0.4 K and 1.19 K, meaning that our measurements at 0.4 K represent the ground state. By fitting the spectrum with a central peak and two equidistant satellites we determine five different internal fields which are listed in Table~\ref{tab:11B_internal_fields}. We determine the peak multiplicity by rounding the scaled peak intensity at 1.19 K to integer values. We observe five peaks, but the sum of the multiplicities is six as Site 2 has about twice the intensity of the other peaks. The order parameter, taken to be the difference between the highest and the lowest local field at a given temperature, decreases only by 0.5 \% from 0.4~K to 1.19~K. 

\begin{figure}[h!]
    \centering
    \includegraphics[width=0.9\textwidth]{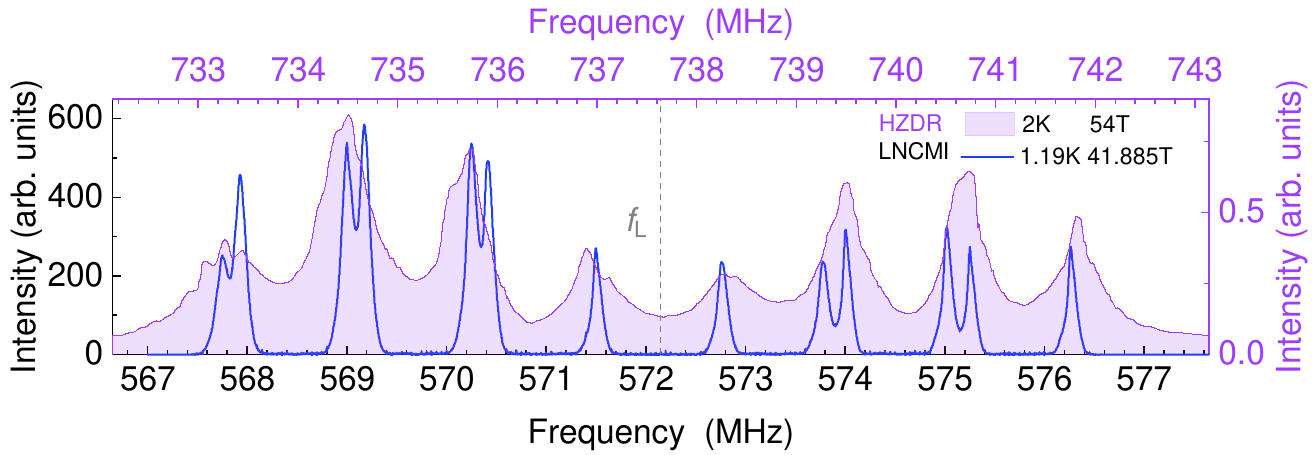}
    \caption{Comparison of $^{11}$B-NMR in the 1/3 plateau at 41.885~T, at the beginning of the plateau, and the pulsed field spectrum measured at 54~T deep in the plateau~\cite{kohlrautz2016}. $f_L$ marks the Larmor frequency for both spectra. Differences in peak widths likely arise due to less stable field in pulsed field conditions. 
    }
    \label{HZDR}
\end{figure}

Earlier work relied on field sweeps in static or pulsed fields~\cite{kohlrautz2016}. In Fig.~\ref{HZDR} we compare our spectrum measured at 41.885~T, at the very edge of the 1/3 plateau, with a pulsed field spectrum at 54~T, deep in the 1/3 plateau. We find that the pulsed field spectrum is only slightly wider. However, the main spectral features are the same. 

\begin{table}[h]
\centering
\begin{tabular}{|c|c|c|c|c|c|c|}
\hline
\textbf{Site} &  \textbf{$H_i$ (T) 0.4 K} & \textbf{$w_i$ (mT) 0.4 K} & \textbf{$H_i$ (T) 1.19 K} & \textbf{$w_i$ (mT) 1.19 K} & \textbf{intensity (arb. units) 1.19 K} & \textbf{Multiplicity} \\
\hline
1 & -0.2096 & 12.48 & -0.2092 & 11.37 & 1.19 & 1 \\
2 & -0.1962 & 9.73 & -0.1953 & 9.58 & 1.97 & 2\\
3 & -0.1172 & 7.58 & -0.1163 & 7.20 & 0.92 & 1\\
4 & 0.1602 & 9.6 & 0.1586 & 8.47 & 0.88 & 1\\
5 & 0.2346 & 10.51 & 0.2328 & 9.39 & 1 & 1\\
\hline
\end{tabular}
\caption{Internal fields, $H_i$, corrected by the demagnetizing field, $H_d$=-0.022 T, Gaussian widths, $w_i$, and intensities from $^{11}$B-NMR in the 1/3 plateau at 41.855 T and at the temperatures 0.4 K and 1.19 K.}
\label{tab:11B_internal_fields}
\end{table}

In a similar way as for the $^{63,65}$Cu spectra, we observe at least 5 different $^{11}$B NMR local fields. As the $^{63,65}$Cu spectra probe dominantly the onsite spin, their \textit{small} line splitting tells us that the local spin polarizations are weakly affected by the 3D coupling. In contrast to that, $^{11}$B NMR lines are strongly separated, meaning that their splitting is dominantly induced by the effects of the direct dipolar coupling to the electronic spins in the adjacent planes. Following the same logic as for Cu NMR, ABAB stacking leads to larger splittings for two out of three boron sites. A large splitting will naturally occur since strongly polarized triplets generate a larger inter-layer dipolar field than the singlet-like wings. Note that both ABAB and ABC stacking of the parallel spin-polarization ``stripes'' [of the in-plane structure; see Fig.~1(g)] reduces the inter-layer repulsion of triplets (see Fig.~\ref{stacking}).

Building on the methods in~\cite{Takigawa2013}, the local fields at the boron sites in the $1/3$ plateau can be analyzed semi-quantitatively as follows.

\begin{figure}[h!]
    \centering
    \includegraphics[width=0.9\textwidth]{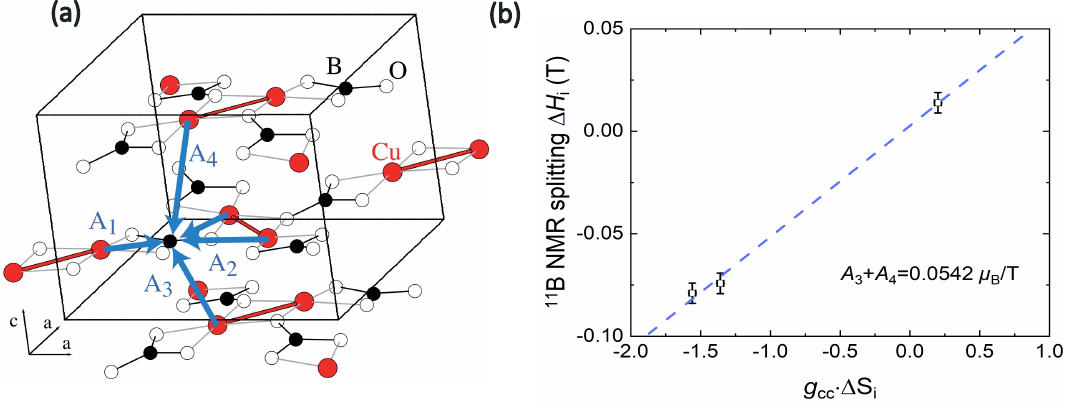}
    \caption{ (a) The crystal structure of \scbo. The arrows indicate the hyperfine coupling of a boron site to the nearest and the second
nearest Cu on the same layer ($A_1$ and $A_2$) as well as the coupling to the third and fourth neighbors on the adjacent layers ($A_3$ and $A_4$), as reproduced from Ref.~\cite{Takigawa2013}. (b) Line splittings $\Delta H_i$ of $^{11}B$ NMR in the 1/3 plateau plotted vs. the differences of the calculated spin polarizations, $g_{cc}\Delta S_i$. The slope corresponds to the inter-layer hyperfine coupling $A_3 + A_4 = 0.0542\,\text{T}/\mu_B$. Vertical error bars are estimated based on the peak widths in table~\ref{tab:11B_internal_fields}, horizontal error bars are estimated from observed peak splittings in Cu NMR in the 1/3 plateau.}
    \label{hyperfine}
\end{figure}

We consider four types of hyperfine coupling ($A_1$, $A_2$, $A_3$, $A_4$), of which $A_1$ and $A_2$ ($A_3$ and $A_4$) are the intra-layer (inter-layer) coupling constants (see the Fig.~\ref{hyperfine}(a). There are six boron sites in the unit cell, therefore six local field values ($H_{0A}$, $H_{0B}$, $H_{1A}$, $H_{1B}$, $H_{2A}$, $H_{2B}$). A and B distinguish different inter-layer configurations. These local fields are expressed as
 \begin{align}\label{fields}
\begin{split}
H_{0A} &= S_0 A_1 + (S_1 + S_2) A_2 + S_0 (A_3 + A_4) \\
H_{0B} &= S_0 A_1 + (S_1 + S_2) A_2 + S_1 (A_3 + A_4) \\
H_{1A} &= S_1 A_1 + (S_1 + S_2) A_2 + S_2 (A_3 + A_4) \\
H_{1B} &= S_1 A_1 + (S_1 + S_2) A_2 + S_0 (A_3 + A_4) \\
H_{2A} &= S_2 A_1 + 2 S_0 A_2 + S_2 (A_3 + A_4) \\
H_{2B} &= S_2 A_1 + 2 S_0 A_2 + S_1 (A_3 + A_4).
\end{split}
\end{align}
The values of line splittings, $\Delta H_i$, due to inter-layer coupling are thus given as 
\begin{align}\label{splitting}
\begin{split}
H_{0A} - H_{0B} &= (S_0 - S_1) (A_3 + A_4) \\
H_{1A} - H_{1B} &= (S_2 - S_0) (A_3 + A_4) \\
H_{2A} - H_{2B} &= (S_2 - S_1) (A_3 + A_4).
\end{split}
\end{align}

The three equations (\ref{splitting}) define a linear dependence of the line splittings $\Delta H_i$ vs. the differences of the calculated spin polarizations, $g_{cc}\Delta S_i$, as shown in Fig.~\ref{hyperfine}(b). The slope corresponds to the thus experimentally determined inter-layer hyperfine couplings $A_3 + A_4 = 0.0542\,\text{T}/\mu_B$, not far from the dipolar coupling 0.074$\,\text{T}/\mu_B$ for point dipoles. This confirms that all boron sites are split and that Site 2 must have a multiplicity of 2 due to an accidental overlap.

In order to determine the intra-layer couplings $A_1$ and $A_2$, we fit the above six equations (\ref{fields}) to the line positions, as shown in Fig.~\ref{11B_NMR}(c), while keeping the inter-layer coupling fixed at $ 0.0542\,\text{T}/\mu_B$ and the average coupling constrained by the ``sum rule'' $A_1+2 A_2+A_3+A_4 = (\Sigma_i H_i) / (g_{cc} \mu_B \cdot 2 \Sigma_i S_i)$.

We find $A_1 = -0.245\,\text{T}/\mu_B$ and $A_2 = 0.0249\,\text{T}/\mu_B$. Note that in our model the dipolar couplings to distant neighbors are neglected, yet we obtain a good fit. Going beyond this approximation will be the subject of subsequent investigations.

\FloatBarrier